\PassOptionsToPackage{unicode}{hyperref}
\PassOptionsToPackage{hyphens}{url}
\PassOptionsToPackage{dvipsnames,svgnames,x11names}{xcolor}
\documentclass[
  11pt,
]{article}
\usepackage{xcolor}
\usepackage[margin=1.1in]{geometry}
\usepackage{amsmath,amssymb}
\usepackage{iftex}
\ifPDFTeX
  \usepackage[T1]{fontenc}
  \usepackage[utf8]{inputenc}
  \usepackage{textcomp} 
\else 
  \usepackage{unicode-math} 
  \defaultfontfeatures{Scale=MatchLowercase}
  \defaultfontfeatures[\rmfamily]{Ligatures=TeX,Scale=1}
\fi
\usepackage{lmodern}
\ifPDFTeX\else
\fi
\IfFileExists{upquote.sty}{\usepackage{upquote}}{}
\IfFileExists{microtype.sty}{
  \usepackage[]{microtype}
  \UseMicrotypeSet[protrusion]{basicmath} 
}{}
\makeatletter
\@ifundefined{KOMAClassName}{
  \IfFileExists{parskip.sty}{%
    \usepackage{parskip}
  }{
    \setlength{\parindent}{0pt}
    \setlength{\parskip}{6pt plus 2pt minus 1pt}}
}{
  \KOMAoptions{parskip=half}}
\makeatother
\usepackage{longtable,booktabs,array}
\usepackage{calc} 
\usepackage{etoolbox}
\makeatletter
\patchcmd\longtable{\par}{\if@noskipsec\mbox{}\fi\par}{}{}
\makeatother
\IfFileExists{footnotehyper.sty}{\usepackage{footnotehyper}}{\usepackage{footnote}}
\makesavenoteenv{longtable}
\providecommand{\tightlist}{%
  \setlength{\itemsep}{0pt}\setlength{\parskip}{0pt}}
\usepackage{bookmark}
\IfFileExists{xurl.sty}{\usepackage{xurl}}{} 
\hypersetup{
  colorlinks=true,
  linkcolor={Maroon},
  filecolor={Maroon},
  citecolor={Blue},
  urlcolor={Blue},
  pdfcreator={LaTeX via pandoc}}

\author{}
\date{}

\begin{document}

\section{Adversarial Test-Hardening for AI-Written Code: An Instrument
Autopsy and a Pre-Registered Causal Estimate of the Critic
Loop}\label{adversarial-test-hardening-for-ai-written-code-an-instrument-autopsy-and-a-pre-registered-causal-estimate-of-the-critic-loop}

\textbf{Jeff Otterson} (W. P. Carey School of Business, Arizona State
University)

\begin{quote}
Disclosure: This manuscript was drafted by an AI assistant from the
author's frozen, pre-registered experimental artifacts under the
author's direction; every quantitative and design claim was checked
against the committed sources. The tooling is disclosed rather than
hidden, consistent with the paper's subject (see Section 3.9).
\end{quote}

\begin{center}\rule{0.5\linewidth}{0.5pt}\end{center}

\subsection{Abstract}\label{abstract}

Large language models increasingly write both code and the tests meant
to check it, and coverage, the usual measure of a suite, records what
ran rather than what was verified. We study an adversarial
test-hardening loop under a mechanical oracle: a Tester model writes
tests, mutation testing names the surviving injected defects, and a
Critic model writes tests to kill exactly those, with every verdict
decided mechanically so that no model judges another model's output. We
report two pre-registered experiments and the instrument autopsy that
connects them. In Experiment 1, on five Python subjects (one
same-lineage-loop cell could not be scored), the loop killed 105 mutants
that one-shot generation missed and lost none, and the
cross-lineage-Critic question (H2) returned a pre-declared null; the
central finding was an autopsy: an earlier analysis had reported a
cross-lineage effect at p = 9.5 x 10\^{}-66 that was a
lineage-correlated instrument artifact, an output cap that silently
truncated the verbose model, caught only by adversarial review of the
completed analysis. A subsequent cross-model review of this manuscript
identified a remaining confound: each arm resampled its own initial test
suite, so Experiment 1's comparisons are between independently sampled
pipelines. Experiment 2 removes it. Under a publicly pre-registered
frozen-shared-round-0 design (five replicates on each of four subjects,
every seed committed before its continuations), same-lineage Critic
rounds killed 78\% of the survivors the frozen initial suite left
standing (mean incremental kill rate 0.783, 95\% cluster-bootstrap
interval {[}0.592, 0.935{]}), a within-replicate causal estimate; and
the cross-provider critic configuration, which Experiment 1's design
could not distinguish from its same-provider control, showed a positive
pilot difference (within-replicate rate gap 0.178, 95\% bootstrap
interval {[}0.039, 0.347{]}; direction stable under every leave-one-out
sensitivity, magnitude dominated by a single replicate) at 5.5x lower
arm cost. This is a comparison of two named model-provider-harness
configurations, not an isolated lineage effect: part of the measured gap
is one configuration's receipted operational failures, including
recurrences of the autopsy's truncation mechanism under the corrected
instrument, this time detected and scored rather than laundered. We
conclude that cross-model comparisons can inherit the asymmetries of the
harness that runs them, that adversarial cross-model review catches what
mechanical gates miss at the instrument, analysis, and manuscript levels
alike, and we release both protocols, all receipts, and the analysis
code.

\begin{center}\rule{0.5\linewidth}{0.5pt}\end{center}

\subsection{1. Introduction}\label{introduction}

Large language models now write both production code and the tests meant
to check it, and the second job is the weaker link. Coverage, the usual
measure of a test suite, records which lines ran, not whether anything
about their behavior was actually verified, and a model asked to write
tests for its own code tends to take the path of least resistance: it
exercises the happy path and asserts weakly, producing suites that are
green and nearly empty of fault-detecting power. A test suite can be
fully covered and catch almost nothing. As models take over more of the
testing work, the evidentiary value of ``the tests pass'' is arguably
falling, precisely where practitioners are relying on it more.

Mutation testing is the standard way to measure a suite's
fault-detecting power directly. It injects small, deliberate defects
(mutants) into the code and counts how many the suite catches: a killed
mutant is a defect the tests detected, and a surviving mutant is a
specific, named gap the tests would let through. We use mutation testing
as a mechanical oracle inside a two-role loop. A Tester model writes an
initial suite; the surviving mutants are handed, by name, to a Critic
model that writes tests to kill exactly those; the loop repeats until no
new mutants die. Every verdict in this loop is mechanical: a generated
test either kills a mutant under the test runner or it does not, and no
model ever judges its own or another model's output. The design
deliberately removes model judgment from the verdict, which is the
channel through which self-preference bias corrupts model-graded
evaluation. Removing model judgment from the verdict is necessary but,
as this study's own experience shows (Section 5), not sufficient to make
a cross-model comparison neutral.

We do not claim this loop is new. Feeding survivor mutants back into a
model prompt and running an adversarial test-versus-mutant loop under a
mechanical oracle are both established in prior work. Our study is a
pre-registered empirical evaluation, and its value is in the discipline
with which it was run and reported, not in the idea. We froze the
hypotheses, the subjects, the measurement scopes, and the success
criteria before any paid run, enforced that freeze in tooling, and
required every reported number to be recomputable from committed per-run
receipts. We test two hypotheses on five pinned Python subjects under
the mutation-kill oracle: H1, that the adversarial loop kills more
mutants than one-shot generation; and H2, that a cross-lineage Critic (a
different-provider model) kills more of the survivors a same-lineage
Tester missed than a same-lineage Critic does. We report cost per killed
mutant throughout.

In Experiment 1, H1 is supported as a comparison of two independently
sampled pipelines: across the paired mutants where the two approaches
disagreed, the loop killed 105 that one-shot generation missed and lost
none, a result that grew stronger, not weaker, when a measurement defect
working against it was corrected. H2 is not supported there; it is a
pre-declared null in a design that, as Experiment 2 shows, could not
have distinguished the configurations. Because each Experiment 1 arm
drew its own initial test suite, those comparisons do not isolate the
loop or the critic lineage from initial-draw sampling variance, a
limitation stated where the arms are defined (Section 3.2).

The paper's central contribution is neither of those results. An earlier
version of the H2 analysis reported an overwhelming cross-lineage effect
(pooled p = 9.5 x 10\^{}-66) that was almost entirely an instrument
artifact: an undetected output ceiling silently truncating the verbose
model's replies, correlated with exactly the variable under test.
Section 4.4 reports that incident once, with receipts, and Section 5.2
interprets it once. The general lesson is that a cross-model comparison
can inherit the asymmetries of the harness that runs it: closing the
model-judgment channel does not make the comparison neutral if the
apparatus feeding the two models is not itself symmetric.

The confound was not caught by us. After Experiment 1 was analyzed and
this manuscript drafted, a cross-model adversarial review of the
manuscript itself identified the round-0 resampling confound as blocking
any causal reading of H1. Rather than reframe the claim, we removed the
confound: Experiment 2 was pre-registered and pushed publicly before its
runner existed, freezing one Tester round-0 suite per replicate (five
per subject, four subjects) and running all three continuations from
that identical state, which makes the loop's incremental effect a
within-replicate paired quantity. The same-lineage Critic rounds killed
a mean 78\% of the survivors the frozen round 0 left standing
(incremental kill rate 0.783, 95\% cluster-bootstrap interval {[}0.592,
0.935{]}); the cross-provider configuration showed a positive pilot
difference (within-replicate rate gap 0.178 {[}0.039, 0.347{]}) at 5.5x
lower arm cost; and the autopsy's mechanism recurred three times under
the corrected instrument, this time detected, receipted, and scored
honestly (Section 4.8).

This paper makes three contributions:

\begin{enumerate}
\def\labelenumi{\arabic{enumi}.}
\tightlist
\item
  A causal, within-replicate estimate of the incremental effect of a
  guarded, add-only Critic continuation over a frozen initial suite,
  under a publicly pre-registered frozen-shared-round-0 design
  (Experiment 2), together with the descriptive pipeline-level
  comparison that preceded it (Experiment 1), with cost per killed
  mutant disclosed throughout, a normalization not found in the studies
  reviewed here.
\item
  A pilot comparison of a cross-provider against a same-provider Critic
  configuration under a mechanical mutation-kill oracle, a comparison we
  could not find in the studies reviewed here: indistinguishable under
  Experiment 1's independently-sampled design, a positive
  within-replicate difference under Experiment 2's paired design,
  reported with per-replicate values and leave-one-out sensitivities,
  and explicitly a configuration comparison (model, provider API
  behavior, and harness interaction bundled), never an isolated lineage
  effect.
\item
  As the durable contribution, a fully receipted autopsy of a
  provider-correlated instrument artifact, its fail-closed repair, and
  the mechanism's detected recurrences under the corrected instrument;
  together with the public artifact (both protocols, all receipts and
  frozen seeds, the deviations log, and the analysis scripts) and the
  outside-the-family review process through which each of these defects
  was caught.
\end{enumerate}

\begin{center}\rule{0.5\linewidth}{0.5pt}\end{center}

\subsection{2. Related Work}\label{related-work}

\subsubsection{2.1 LLM-based test
generation}\label{llm-based-test-generation}

A large recent literature uses large language models to generate unit
tests directly. TestPilot {[}arXiv:2302.06527, TSE 2024{]} generates
JavaScript tests from signatures and documentation usage examples with
no additional training; ChatTester {[}arXiv:2305.04207{]} and
ChatUniTest {[}arXiv:2305.04764{]} add generate-then-repair loops that
fix compilation and assertion errors in model-written tests. For Python
specifically, CoverUp {[}arXiv:2403.16218, FSE 2025{]} interleaves
coverage analysis with iterative model dialogs to target uncovered
lines, and CodaMosa {[}ICSE 2023{]}, building on the search-based
generator Pynguin (Lukasczyk and Fraser), queries a model to re-seed
search when coverage plateaus. HITS {[}arXiv:2408.11324, ASE 2024{]}
decomposes complex methods into slices for higher coverage. At
industrial scale, TestGen-LLM {[}arXiv:2402.09171, FSE 2024{]} extends
existing test classes only when a generated test verifiably builds,
passes reliably, and increases coverage, using that filter to eliminate
hallucinated tests. Adjacent lines of work target the oracle directly,
generating assertions rather than whole suites (TOGA
{[}arXiv:2109.09262, ICSE 2022{]}), or use models as fuzzers against
crash and inconsistency oracles rather than regression suites (TitanFuzz
{[}arXiv:2212.14834, ISSTA 2023{]}; Fuzz4All {[}arXiv:2308.04748, ICSE
2024{]}); we exclude fuzzing from the comparison because its oracle
(crashes, differential behavior) and artifact (inputs, not persistent
test suites) differ from ours. These systems establish that models can
write useful tests and that a filter is needed to trust them, but they
gate on coverage or on compile-and-pass repair, not on fault detection.
Our work differs in the gate: it scores generated tests by the mutants
they kill, a stronger signal than coverage, and it is the adversarial
survivor-feedback loop, not one-shot generation, that our H1 evaluates.
This study does not include a non-LLM search-based baseline such as
Pynguin or EvoSuite; the comparison of interest here is one-shot versus
adversarial-loop LLM generation, not LLM versus search-based generation.

\subsubsection{2.2 Mutation testing with language
models}\label{mutation-testing-with-language-models}

Mutation testing supplies the mechanical oracle for this work. Two
surveys frame what that oracle can and cannot claim. The oracle problem
{[}Barr et al., IEEE TSE 2015{]} is the general difficulty of deciding
what a test should assert; our pass-on-pristine validation plus
mutant-kill scoring is a partial oracle, and the wrong-oracle tests we
prune (generated tests that fail on correct code) are its concrete
residue. Papadakis et al.'s mutation-testing survey {[}Advances in
Computers 2019{]} treats equivalent mutants, mutation-score
interpretation, and the non-independence of mutants, all three of which
appear in this study as disclosed limitations rather than solved
problems. Several systems combine mutation testing with language models.
MuTAP {[}arXiv:2308.16557; IST 2024{]} re-prompts a single model with
its surviving mutants (produced by the rule-based tool MutPy) to
generate tests that kill them, halting when no survivors remain; it is
single-agent, benchmark-scale, and reports mutation score with no dollar
cost. Meta ACH {[}arXiv:2501.12862, FSE 2025{]} uses three agents (fault
generation, equivalence detection, and test generation) with a single
backbone model to produce few, concern-specific mutants and hardening
tests at industrial scale, under an assured generate-and-test filter.
LLMorpheus {[}arXiv:2404.09952{]} instead generates the mutants
themselves with a model, producing diverse bug-like mutants beyond a
standard operator set, and a recent empirical study {[}arXiv:2406.09843,
TOSEM{]} compares such model-generated mutants against real bugs at
scale. Our mutants are rule-based (mutmut), as in MuTAP, rather than
model-generated as in LLMorpheus and AdverTest; this keeps the injected
defect independent of any model under test.

\subsubsection{2.3 Adversarial and dual-agent test-versus-mutant
loops}\label{adversarial-and-dual-agent-test-versus-mutant-loops}

The closest neighbor is AdverTest {[}arXiv:2602.08146{]}, which runs a
genuine dual-agent adversarial loop: a test agent and a mutant agent
iteratively attack each other under an execution-based mutation-kill
oracle, with measured fault-detection gains over prior baselines.
Crucially, both of AdverTest's agents share a single backbone model per
configuration, and no cross-lineage configuration is measured. Our loop
replicates this adversarial-under-a-mechanical-oracle pattern, and we
claim it as a replication, not an invention (Section 5.1). What
AdverTest and the survivor-feedback lineage of MuTAP do not do, and what
our H2 compares, is a cross-provider critic configuration against a
same-provider control under the mechanical oracle (a comparison of the
two named configurations; Section 3.2).

\subsubsection{2.4 Cross-model diversity, self-preference, and
evaluation
bias}\label{cross-model-diversity-self-preference-and-evaluation-bias}

The premise that a mechanical oracle is worth its cost rests on known
failures of model-graded evaluation: LLM-as-judge protocols carry
position, verbosity, and self-enhancement biases {[}Zheng et al.,
arXiv:2306.05685, NeurIPS 2023{]}, and large language model evaluators
can recognize and favor their own generations {[}Panickssery, Bowman,
and Feng, arXiv:2404.13076, NeurIPS 2024{]}, a self-preference bias a
model-graded verdict cannot escape. Heterogeneous-model methods such as
multi-agent debate {[}Du et al., arXiv:2305.14325, ICML 2024{]} motivate
the intuition that mixing model lineages helps, but the direction is
genuinely open: Self-MoA reports same-model repeated sampling beating
heterogeneous mixtures on reasoning benchmarks, and Refute-or-Promote
{[}arXiv:2604.19049{]} argues a cross-model critic catches correlated
blind spots but does so in defect discovery, without a controlled
same-lineage ablation and without a purely mechanical oracle. Most
directly relevant to our Discussion, recent work argues that models
increasingly share the same errors, which undermines the reliability of
using one model to oversee another {[}Great Models Think Alike and this
Undermines AI Oversight, arXiv:2502.04313{]}. Our instrument autopsy
(Section 5) is a concrete, receipted instance of a related but distinct
hazard: even with model judgment removed from the verdict entirely, an
asymmetry in the harness that feeds the two models can manufacture a
large false cross-model difference.

\subsubsection{2.5 Cost accounting and
pre-registration}\label{cost-accounting-and-pre-registration}

Where cost is reported in this literature, it is normalized to an input
unit rather than an outcome. AdverTest reports an average dollar cost
per method-generation run (for example \$0.270 on Defects4J), and
TestForge {[}arXiv:2503.14713{]} reports dollars per file and per
iteration, but neither divides cost by mutants killed or faults
detected. The nearest per-outcome figure is in an adjacent domain,
defect discovery, where a cross-model critic system reports a
per-discovered-vulnerability cost. We report cost per mutation kill, a
normalization not found in the studies reviewed here. Separately,
pre-registered and registered-report protocols have precedent in
software engineering {[}arXiv:2302.03649{]} and have begun to wrap
LLM-for-software-engineering studies {[}arXiv:2606.10702{]}, though they
remain uncommon in this literature; we adopt pre-registration as a
matter of rigor and claim no priority for it.

\subsubsection{2.6 Positioning}\label{positioning}

Among the studies reviewed here, none measures the effect of
critic-generator lineage or provider diversity on test-suite fault
detection under a mechanical mutation-kill oracle, and none reports LLM
test-generation cost normalized per killed mutant. We do not claim to be
first to run an adversarial test-versus-mutant loop under a mechanical
oracle (AdverTest does this), first to feed survivor mutants into model
prompts (MuTAP does this), or first to pre-register a
software-engineering protocol. We claim a paired cross-provider versus
same-provider critic-configuration comparison under a mechanical oracle
(a configuration comparison, never an isolated lineage effect), a
per-outcome cost normalization not found in the studies reviewed here, a
causal within-replicate estimate of the Critic continuation's
incremental effect together with a descriptive pipeline-level comparison
in the MuTAP-to-AdverTest direction, and, as the durable contribution, a
fully receipted autopsy of a provider-correlated instrument artifact.
The broader literatures on experimental-infrastructure validation and
execution-environment nondeterminism are engaged here only where they
touch this study's own defects; we do not survey them.

\begin{center}\rule{0.5\linewidth}{0.5pt}\end{center}

\subsection{3. Method}\label{method}

\subsubsection{3.1 The adversarial test-hardening
loop}\label{the-adversarial-test-hardening-loop}

We evaluate a two-role loop over a mechanical mutation-kill oracle. A
Tester model is prompted once to write a pytest suite for a target
module (round 0). The suite is run under mutation testing (mutmut): each
surviving mutant is an injected defect the suite failed to catch. A
Critic model is then handed the named survivors, the specific mutants
left alive together with their diffs, and prompted to write tests that
kill exactly those. The loop repeats. After each Critic round the suite
is re-measured, newly killed survivors are removed from the target list,
and the next round attacks what remains. A run stops when two
consecutive rounds kill nothing new (dry\_rounds = 2) or after four
rounds (max\_rounds = 4), whichever comes first.

The verdict at the core of this loop is mechanical: a generated test
either kills a mutant under pytest or it does not, and no model ever
judges its own or another model's output. This closes the
self-preference channel that undermines LLM-as-judge evaluations, where
a model scoring model output has no ground truth. We are careful not to
overstate what this buys. The mechanical oracle removes model judgment
from the verdict itself; it does not make the surrounding measurement
apparatus neutral. Section 5 documents a case where the harness around
the oracle opened a lineage-correlated channel into the result even
though the oracle itself never rendered a subjective judgment.

We do not claim the loop is novel. Feeding survivor mutants into an LLM
prompt was established by MuTAP, and a dual-agent adversarial
test-versus-mutant loop under a mechanical oracle was established by
AdverTest. Our contribution is a controlled measurement, described
below, and a fully receipted account of what it took to make that
measurement trustworthy (Section 5).

\subsubsection{3.2 Arms and hypotheses}\label{arms-and-hypotheses}

Each subject module is run through three arms, all sharing one fixed
Tester model (claude-sonnet-5, Anthropic) so that the only deliberate
variable is the Critic:

\begin{itemize}
\tightlist
\item
  oneshot: Tester round 0 only, no Critic round. This is the one-shot
  baseline.
\item
  loop-same: Tester round 0, then Critic rounds where the Critic is the
  same lineage as the Tester (claude-sonnet-5, Anthropic).
\item
  loop-cross: Tester round 0, then Critic rounds where the Critic is a
  different lineage (gpt-5.6-terra, OpenAI).
\end{itemize}

The cross-lineage Critic is deliberately price-matched to the
same-lineage Critic (posted token rates \$2.50/\$15 per million versus
\$3/\$15) rather than paired with the flagship variant, so that the
comparison is not additionally confounded with price tier; we have no
independent evidence of capability-tier equivalence across providers and
claim none. Price matching cannot isolate lineage: the arms compare two
named model-provider-harness configurations, and ``cross-lineage'' is
used throughout as the descriptive label of the arms, never as the
inferred cause of a difference. We pin real model identifiers
throughout, rather than anonymizing, because the cost accounting is tied
to specific rate cards and the instrument finding in Section 5 is
provider-specific and cannot be stated without them.

Two hypotheses follow:

\begin{itemize}
\tightlist
\item
  H1 (replication): the adversarial loop (loop-same) kills more mutants
  than one-shot generation (oneshot), holding the Tester model fixed.
\item
  H2 (pre-registered wording): a cross-lineage Critic (loop-cross) kills
  more of the survivors the same-lineage Tester missed than a
  same-lineage Critic (loop-same) does. ``Cross-lineage'' is the
  pre-registered arm label; what the arms instantiate are the two named
  model-provider-harness configurations above, and every H2 statement in
  this paper is a claim about those configurations.
\end{itemize}

H1 is framed strictly as a replication of the MuTAP-to-AdverTest loop
direction in a new agentic, repo-level, Python setting with disclosed
per-outcome costs. H2 is the claim we could not find measured in the
studies reviewed here: none measures the effect of critic-configuration
diversity (cross-provider versus same-provider) on test-suite fault
detection under a mechanical mutation-kill oracle. Both directions of
the H2 result were pre-declared publishable, including a null, so the
direction was not assumed in our favor.

One design limitation is inherent to these arms and we state it where
the arms are defined rather than burying it later. Each arm
independently regenerates its own round 0, so the tester suite and its
resulting survivor set are not held fixed between loop-same and
loop-cross. The two critic arms therefore differ in both the round-0
tester draw and the critic lineage, and single-run tester draws carry
substantial variance (Section 6 reports the same tester producing
round-0 kill counts of 69, 168, 50, 62, and 65 across repeated attempts
on one subject). A cleaner design freezes one round-0 survivor set and
runs every continuation from that identical state. Experiment 1 did not,
so its H1 and H2 estimates are comparisons of independently sampled
pipelines, and any small cross-lineage difference it reports must be
read against that confound. An external cross-model review of this
manuscript subsequently identified exactly this as blocking a causal
reading, and Experiment 2 (Section 3.10) is the frozen-round-0 design
that removes it.

\subsubsection{3.3 Subjects and selection}\label{subjects-and-selection}

Five Python subjects, each pinned by commit SHA, each contributing one
paired subject-module cell to every arm. Subjects were selected against
criteria fixed before selection (recorded in the pre-registration's
selection log). The two third-party subjects were drawn by scanning the
most-downloaded PyPI packages in rank order and taking the first two
that met every pre-registered criterion with no discretion: pure Python
(no C extensions), a permissive license, an existing pytest suite (which
we strip in the clone), a plain-logic module of 100 to 800 source lines
(no network- or IO-heavy modules; the first qualifying module in
alphabetical order becomes the target), at least 40 mutants generated by
mutmut on that module, and no prior authorship, contribution, or
analysis by the author. The three author-repository subjects are a
convenience sample chosen for thesis relevance, including the degenerate
zero-kill baseline case described below. This convenience component is a
disclosed threat to external validity (Section 6), compounded by
training-data contamination since all five are public code the models
may have seen in pretraining. For each subject exactly one target module
is mutated, and sibling files needed for imports are carried into the
mutation sandbox unmutated. One subject (graph-guard) was pre-declared
as a pilot with a go/no-go gate before the full grid, and its pilot
cells were pre-declared to count as data; in the event those cells were
later reclassified for an instrument defect (Section 3.8) and rerun, so
the pilot designation does not affect the counted dataset.

{\def\LTcaptype{none} 
\begin{longtable}[]{@{}
  >{\raggedright\arraybackslash}p{(\linewidth - 10\tabcolsep) * \real{0.1000}}
  >{\raggedright\arraybackslash}p{(\linewidth - 10\tabcolsep) * \real{0.1750}}
  >{\raggedright\arraybackslash}p{(\linewidth - 10\tabcolsep) * \real{0.4000}}
  >{\raggedleft\arraybackslash}p{(\linewidth - 10\tabcolsep) * \real{0.0625}}
  >{\raggedleft\arraybackslash}p{(\linewidth - 10\tabcolsep) * \real{0.1250}}
  >{\raggedleft\arraybackslash}p{(\linewidth - 10\tabcolsep) * \real{0.1375}}@{}}
\toprule\noalign{}
\begin{minipage}[b]{\linewidth}\raggedright
Subject
\end{minipage} & \begin{minipage}[b]{\linewidth}\raggedright
Provenance; suite
\end{minipage} & \begin{minipage}[b]{\linewidth}\raggedright
Target module
\end{minipage} & \begin{minipage}[b]{\linewidth}\raggedleft
Mutants
\end{minipage} & \begin{minipage}[b]{\linewidth}\raggedleft
Killed by existing suite
\end{minipage} & \begin{minipage}[b]{\linewidth}\raggedleft
Baseline survivors (scored)
\end{minipage} \\
\midrule\noalign{}
\endhead
\bottomrule\noalign{}
\endlastfoot
attrition-risk-ml & author repo (public); kept & \texttt{src/train.py} &
255 & 0 & 255 \\
graph-guard & author repo (public); kept & \texttt{graph\_guard/ppr.py}
& 80 & 58 & 22 \\
rag-guard & author repo (public); kept & \texttt{rag\_guard/guard.py} &
71 & 46 & 25 \\
packaging & pypa/packaging (OSS); stripped &
\texttt{src/packaging/\_elffile.py} & 69 & 0 & 69 \\
idna & kjd/idna (OSS); stripped & \texttt{idna/cli.py} & 187 & 0 &
187 \\
\end{longtable}
}

Both mutation denominators are shown, per the pre-registration's rule
that neither the full-mutant count nor the baseline-survivor count is
presented alone. The two third-party subjects have their existing suites
stripped in the local clone before any run, so the loop is scored
against a genuinely empty starting suite and never against the upstream
project's own tests; the three author-repo subjects keep their existing
suites, so their scored denominator is the survivors an already-present
suite left alive.

One subject, attrition-risk-ml, is degenerate by design and kept
deliberately. Its existing suite kills 0 of 255 mutants, because the
target module is never imported by any test, so it is the
maximal-headroom false-pass case: a suite that looks present but catches
nothing, exactly the failure mode this line of work exists to expose.
Because any ratio against a zero-kill baseline is undefined and any
intervention trivially wins against zero, this subject is reported with
absolute kill counts only, never a relative-improvement figure.

We perform no equivalent-mutant detection. Some mutants in each subject
may be semantically equivalent to the original and therefore unkillable
by any test, which means every reported kill rate is a lower bound. This
does not affect the paired comparisons: both arms in any comparison face
the identical mutant set, so an unkillable mutant simply lands in the
both-miss cell that McNemar discards.

\subsubsection{3.4 The mechanical oracle}\label{the-mechanical-oracle}

Mutation testing is run with mutmut, which injects rule-based,
single-change mutants (one mutation per variant), in contrast to
LLM-generated mutants used by some adversarial-loop systems. Each
subject's mutation scope, meaning the exact module mutated and the exact
sibling files copied into the sandbox, was fixed at pre-registration and
thereafter changed only through logged, receipted amendments (Section
3.8), never silently at run time. Kills are counted per mutant over each
subject's own pristine-baseline survivor set: the survivors measured on
the pristine module before any generated test exists, so both arms in
any comparison are scored against the same universe of mutants. A
generated test file is rejected outright, crediting zero kills with the
rejection and its reason recorded, when it fails validation as a file:
it does not parse, the reply was truncated or malformed, or none of its
tests survive the per-test checks. Within an accepted file, each
individual test must pass on the pristine (un-mutated) module twice (the
flake check); a test that fails on pristine code, a wrong-oracle test
rather than a bad file, is pruned individually and logged, never counted
as a kill, and the round continues with the tests that survive (a file
that passes the pristine run once and fails the repeat is rejected whole
as flaky -- a code path no counted round ever triggered).

\subsubsection{3.5 The paired statistic}\label{the-paired-statistic}

Because every arm runs against the same pinned subject clone, kill
outcomes are paired mutant-for-mutant across arms. The primary statistic
is a two-sided exact McNemar test on the discordant pairs: b = mutants
killed by arm A only, c = mutants killed by arm B only. Pairs where both
arms agree (both kill, or both miss) carry no information and are
excluded by construction; the evidence lives entirely in the
disagreements. The test uses the doubled exact binomial tail (min-tail
doubling, with n = 0 defined as p = 1), not the chi-square
approximation, which is appropriate for the small discordant counts
these subjects produce.

The pooled test treats individual mutants as the unit of analysis, which
assumes independence across mutants; in fact mutants are clustered by
subject (idna alone contributes 64 of the 105 H1 discordant pairs), and
within-subject correlation among mutants of one module is not modeled.
This inflates the nominal pooled significance, which is why we
pre-registered per-subject tables as a companion readout and report them
alongside every pooled figure rather than resting on the pool alone.

Each hypothesis is reported in three pre-declared views, all fixed
before any data. First, one pooled McNemar across all subjects'
discordant pairs, designated the confirmatory statistic. Second, a
pooled view excluding the degenerate subject, whose 255 paired outcomes
would otherwise dominate the pool. Third, per-subject McNemar tables.
All three views were committed in advance and all three are reported for
both hypotheses, so no view can be chosen, promoted, or dropped after
seeing results. We ran no a priori power analysis; with three of four
scorable subjects at or near a 98.6 to 100 percent kill ceiling, the
number of discordant pairs available to H2 was near its structural
minimum, so the H2 null is stated throughout as ``no effect detectable
at this design's power,'' never as ``no effect.''

\subsubsection{3.6 Pre-registration, enforced by
tooling}\label{pre-registration-enforced-by-tooling}

The protocol, meaning the hypotheses, arms, subjects, frozen scopes, the
exact statistic, and the supported-or-not-supported criteria, was
written and committed before any paid run. The freeze is not a promise.
The experiment command mechanically refuses to run any arm unless the
committed machine-readable protocol is byte-identical to the repository
HEAD, and every reported number is required to be recomputable from
committed per-run receipts (the per-round JSON logs recording token
counts, cost, verdicts, and preserved test artifacts for every run). The
success criteria were fixed in advance: H1 is supported if and only if
the pooled McNemar yields p \textless{} 0.05 and the direction favors
loop-same; H2 if and only if the pooled McNemar yields p \textless{}
0.05 and the direction favors loop-cross. Any other outcome, in either
case, is not supported, and a null was pre-declared to be reported with
the same prominence as a positive result. No subgroup analysis was
permitted to substitute for the three pre-declared views; any post-hoc
breakdown is labeled exploratory and never replaces the primary readout.

Each arm runs once per subject under a fixed grid (three arms times five
subjects, fifteen cells). A cell that produces a valid verdict is never
rerun in search of a better one, and there is no early stopping for
significance. A cell that is invalidated or aborts may be rerun only
under a logged deviation. One subject's loop-same cell was attempted
five times; before the final attempt, a stopping rule was pre-committed
in the deviations log (if that attempt also failed, the cell would be
scored missing, with no further gates), and when it failed the cell was
recorded as missing data rather than backfilled or estimated. Every
departure from the frozen protocol is logged in an append-only
deviations file. Fourteen of the fifteen cells produced valid verdicts;
the one missing cell (attrition-risk-ml loop-same) removes the largest
subject from the paired statistics, so all paired analyses span four
subjects, and the pre-declared with-degenerate and without-degenerate
pooled views coincide in composition for this dataset (a composition
collapse we state explicitly).

\subsubsection{3.7 Cost accounting}\label{cost-accounting}

Every model round is metered exactly from its input and output token
counts; an unpriced model fails the run closed rather than pricing at a
guessed rate. We report cost-per-kill, meaning total cost divided by
mutants killed, per arm. This per-outcome normalization is not found in
the studies reviewed here, which report cost per method-run or per file;
the adjacent defect-discovery literature does report a per-outcome
figure (a cross-model critic system reports a dollar cost per discovered
vulnerability), so we scope the novelty to test generation rather than
claiming it broadly.

\subsubsection{3.8 Instrument integrity}\label{instrument-integrity}

The measurement instrument was itself treated as a subject of scrutiny,
and the honest account of what that surfaced falls into three tiers
rather than one.

First, defects caught at zero model cost, before any model was called,
by a model-free dry-run validator built partway through the study:
scope-transcription errors and a mutation-sandbox failure that would
have silently recorded every mutant as unchecked. From the point that
validator existed, and later a canary must-kill probe that confirms the
sandbox actually collects a freshly written test, each subsequent scope
fix was validated by a zero-cost dry run before any further paid cell
ran.

Second, defects caught only after paid cells had already run, which
forced reclassification of already-spent data. A trampoline crash that
laundered into a false all-survived zero, and an include-list scoping
bug that silently prevented freshly generated tests from being
collected, together invalidated several paid cells, including two that
had previously been counted as data. More was ultimately spent on runs
that were invalidated and rerun than on the runs finally counted; we
report the invalidated spend alongside the counted spend rather than
omitting it.

Third, the one defect caught only after all fifteen counted cells had
run, by a final adversarial review of the completed analysis rather than
by any runtime gate. Because it changed the headline result, it is
documented in full as a finding in its own right (Section 5). In every
tier the repair principle was the same: make the instrument fail loudly
rather than record a plausible-looking wrong number. Across the study
the protocol carried ten numbered amendments plus one prompt-template
amendment, each logged with its rationale and its cost consequences.

\subsubsection{3.9 Reproducibility and
disclosure}\label{reproducibility-and-disclosure}

All counted cells ran under a single fixed prompt version, with one
per-subject import hint added for the one subject whose package layout
required it; the prompt texts are content-hashed and the hashes are
recorded in the pre-registration (16-character prefixes: tester
98abbb8532990865, critic 703335c0af111836). The models were called at
provider default sampling: no temperature or top-p value is set anywhere
in the experiment's provider configuration, and the only
sampling-relevant request-body parameter is the output-token cap
discussed in Section 5. Mutation testing used mutmut 3.6.0 (pinned) with
pytest 8 or newer; the package requires Python 3.11 or newer and its
continuous integration validates on Python 3.11, 3.12, and 3.13. All
runs executed on macOS on Apple Silicon, the platform whose fork
behavior produced the missing-cell deadlock analyzed in Section 6. The
protocol, the machine-readable arms configuration, the per-run receipts,
the deviations log, and the analysis script that recomputes the reported
statistics are committed to the public repository, so every reported
statistic is reproducible from the receipts. A small, bounded portion of
the all-in spend total is disclosed as an estimate rather than
receipt-derived, where a crash billed a call before its receipt row
landed. An automated reproduction command is provided. Experiment 2 adds
its own artifacts under the same discipline: its protocol was committed
and pushed to the public repository before its runner existed and before
any paid run (the pushed commit is the pre-registration timestamp; we
note this is a public, hash-chained record rather than a third-party
registration service), the runner was required, as a pre-registered
precondition, to enforce the freeze mechanically before the first paid
call, every frozen seed is committed before its continuations may run
(twenty of the twenty-one seeds were also pushed publicly before any of
their continuations ran; the one mid-run replacement seed was committed
locally before its continuations and pushed with the final receipts, so
commit-before-continuation is the mechanically enforced gate and the
public push is the audit trail), and the analysis script's bootstrap
draw count (10,000) and random seed are frozen in the protocol file
itself, so every reported interval is bit-for-bit reproducible. The
manuscript was drafted by an AI assistant from these frozen artifacts
under the author's direction, and every quantitative and design claim
was checked against the committed sources; this tooling is disclosed
rather than hidden, consistent with the paper's subject. The
pre-registration's limitations (training-data contamination, no syscall
sandbox against mutant-environment gaming, a two-run flake check that a
one-third-flaky test still clears roughly four times in nine, a single
fixed Tester model, one subject's stdin code path, and one subject's
pinned-module divergence from its development branch) are carried into a
dedicated Threats to Validity section rather than asserted away here.

\subsubsection{3.10 Experiment 2: the frozen-shared-round-0 causal
design}\label{experiment-2-the-frozen-shared-round-0-causal-design}

Experiment 2 removes the initial-draw confound with one mechanical
change: the round-0 suite becomes frozen shared state instead of a
per-arm sample. For each replicate, the Tester is called once; the
resulting suite is validated by the standard guardrails, measured once,
and frozen, the test file bytes, the pristine baseline survivor set, and
the post-round-0 survivor set committed to the repository before any
continuation may run (the same committed-byte-identity gate that
enforces the protocol freeze is applied to each seed directory). Three
continuations then run from that identical state: no-critic (the frozen
round 0 is the outcome, at zero marginal cost), loop-same (same-lineage
Critic rounds), and loop-cross (cross-lineage Critic rounds), with the
loop parameters of Experiment 1. The loop's incremental effect is
therefore a within-replicate paired quantity: every continuation of a
replicate starts from byte-identical tests and an identical survivor
set, so arm differences cannot come from initial-draw variance. Five
replicates per subject, drawn independently, average over the round-0
draw distribution itself.

Four of Experiment 1's five subjects are used, under the same pinned
commits and frozen scopes; the degenerate zero-kill-baseline subject is
excluded, pre-declared, both for its unresolved nested-parallelism
deadlock (four failed attempts in Experiment 1) and because its
special-case analysis rules add nothing to a causal estimate. The
protocol pre-registers outcome-blind procedure for every contingency:
seed draws are capped at three per replicate and seven per subject,
counted mechanically from receipts on disk; a draw is accepted or
rejected only by the mechanical validity gates, never by its kill count;
a seeded continuation re-verifies the frozen measurement before any
Critic call (the pristine baseline and the post-injection survivor set
must reproduce exactly), and any mismatch invalidates the whole
replicate, retires the seed in place, and makes a replacement draw
mandatory if the caps permit; a validly scored cell is never rerun; an
aborted or crashed cell is rerun exactly once, then scored missing.
Every one of these rules is enforced by the tooling, not by the
operator, and every departure is logged in the same append-only
deviations file as Experiment 1.

The pre-registered estimands are effect sizes with uncertainty
intervals, not p-value gates, and the causal quantity is deliberately
narrow: the incremental mutation kills produced by this guarded,
add-only Critic continuation, conditional on these frozen seeds,
subjects, models, and stopping rules. Nothing broader, no
cross-repository or cross-model generalization, is identified by the
design. Because generated tests are add-only and both loop continuations
start from the identical frozen suite, a continuation's incremental
kills over the seed (Delta) cannot be negative by construction, so the
pre-registered question is magnitude, never sign. We report, per
replicate, Delta for each loop arm and the incremental kill rate rho =
Delta / \textbar S\textbar{} over the frozen post-round-0 survivor set S
(undefined when \textbar S\textbar{} = 0, an exclusion fixed in
advance), and for the cross-lineage question the within-replicate
differences D = Delta\_cross - Delta\_same and rho\_gap = rho\_cross -
rho\_same. Per subject, each estimand is summarized by its mean over the
five replicates with a 95\% percentile bootstrap interval; the pooled
figure is the unweighted mean of the four subject means with a
subject-level cluster bootstrap, resampling subjects, because replicates
within a subject are correlated and mutants within a replicate far more
so. No pooled per-mutant McNemar is used as a confirmatory statistic in
Experiment 2, for exactly the clustering reason Section 3.5 discloses
about Experiment 1's pooled view. The bootstrap draw count and random
seed are frozen in the protocol file. The cross-lineage estimand is
pre-registered as a pilot: no supported-or-not verdict attaches to it,
because four subjects cannot power a lineage claim and because
``lineage'' here bundles model, provider API behavior, and harness
interaction, a bundling the autopsy makes vivid.

\begin{center}\rule{0.5\linewidth}{0.5pt}\end{center}

\subsection{4. Results}\label{results}

Sections 4.1 through 4.5 report Experiment 1; Sections 4.6 through 4.9
report Experiment 2. Every figure below is recomputed from committed
per-run receipts by the committed analysis scripts, driven for
Experiment 1 by a committed manifest of which runs count and for
Experiment 2 by the frozen seeds and receipts themselves. This analysis
supersedes an earlier one: a final adversarial review of the completed
analysis found a truncation defect sitting inside the counted cells (an
output cap with no truncation detection), which reclassified four
counted cells as instrument-invalid and confounded the entire H2
readout. The version reported here is computed from the corrected
instrument with the affected cells rerun. The headline difference is
itself the paper's central finding: H2's previously significant result
did not survive the correction, and the Discussion analyzes why.

\subsubsection{4.1 Counted cells}\label{counted-cells}

The design is five subjects times three arms, fifteen cells. Fourteen
produced valid verdicts; one, attrition-risk-ml loop-same, is a missing
cell, after its original run was reclassified instrument-invalid and
four rerun attempts each failed on a nested-parallelism deadlock that
survived three targeted, individually verified instrument fixes; the
missing verdict was recorded under a stopping rule pre-committed before
the final attempt. Kill counts are per mutant over each subject's own
pristine-baseline survivor set. The verdict column records how each run
terminated: oneshot (baseline, no critic round); clean (all baseline
survivors killed); dry (two consecutive rounds killed nothing new);
rounds-cap (reached the four-round limit); missing (no valid verdict
within the attempt budget). Dropped counts wrong-oracle tests pruned
across all rounds and never credited as kills.

{\def\LTcaptype{none} 
\begin{longtable}[]{@{}
  >{\raggedright\arraybackslash}p{(\linewidth - 14\tabcolsep) * \real{0.1034}}
  >{\raggedright\arraybackslash}p{(\linewidth - 14\tabcolsep) * \real{0.1034}}
  >{\raggedright\arraybackslash}p{(\linewidth - 14\tabcolsep) * \real{0.1034}}
  >{\raggedleft\arraybackslash}p{(\linewidth - 14\tabcolsep) * \real{0.1379}}
  >{\raggedleft\arraybackslash}p{(\linewidth - 14\tabcolsep) * \real{0.1379}}
  >{\raggedleft\arraybackslash}p{(\linewidth - 14\tabcolsep) * \real{0.1379}}
  >{\raggedleft\arraybackslash}p{(\linewidth - 14\tabcolsep) * \real{0.1379}}
  >{\raggedleft\arraybackslash}p{(\linewidth - 14\tabcolsep) * \real{0.1379}}@{}}
\toprule\noalign{}
\begin{minipage}[b]{\linewidth}\raggedright
Subject
\end{minipage} & \begin{minipage}[b]{\linewidth}\raggedright
Arm
\end{minipage} & \begin{minipage}[b]{\linewidth}\raggedright
Verdict
\end{minipage} & \begin{minipage}[b]{\linewidth}\raggedleft
Baseline survivors
\end{minipage} & \begin{minipage}[b]{\linewidth}\raggedleft
Killed
\end{minipage} & \begin{minipage}[b]{\linewidth}\raggedleft
Rate
\end{minipage} & \begin{minipage}[b]{\linewidth}\raggedleft
Cost (\$)
\end{minipage} & \begin{minipage}[b]{\linewidth}\raggedleft
Dropped
\end{minipage} \\
\midrule\noalign{}
\endhead
\bottomrule\noalign{}
\endlastfoot
graph-guard & oneshot & oneshot & 22 & 12 & 54.5\% & 0.13 & 0 \\
graph-guard & loop-same & dry & 22 & 17 & 77.3\% & 1.02 & 1 \\
graph-guard & loop-cross & clean & 22 & 22 & 100.0\% & 0.35 & 3 \\
rag-guard & oneshot & oneshot & 25 & 21 & 84.0\% & 0.06 & 0 \\
rag-guard & loop-same & clean & 25 & 25 & 100.0\% & 0.11 & 0 \\
rag-guard & loop-cross & clean & 25 & 25 & 100.0\% & 0.16 & 1 \\
packaging & oneshot & oneshot & 69 & 36 & 52.2\% & 0.11 & 0 \\
packaging & loop-same & rounds-cap & 69 & 68 & 98.6\% & 0.64 & 0 \\
packaging & loop-cross & dry & 69 & 68 & 98.6\% & 0.30 & 3 \\
idna & oneshot & oneshot & 187 & 123 & 65.8\% & 0.12 & 0 \\
idna & loop-same & clean & 187 & 187 & 100.0\% & 0.65 & 9 \\
idna & loop-cross & clean & 187 & 187 & 100.0\% & 0.19 & 0 \\
attrition-risk-ml & oneshot & oneshot & 255 & 62 & 24.3\% & 0.07 & 0 \\
attrition-risk-ml & loop-same & missing & 255 & n/a & n/a & 0.07 & 0 \\
attrition-risk-ml & loop-cross & rounds-cap & 255 & 254 & 99.6\% & 0.34
& 2 \\
\end{longtable}
}

Costs are shown to two decimals for readability; the aggregate figures
in Section 4.5 are computed from the full-precision receipts, so summing
the rounded column will not exactly reproduce them. Nineteen
wrong-oracle tests were pruned across the fourteen valid cells (nine of
them in idna loop-same alone), each logged and never counted as a kill.

Because attrition-risk-ml's loop-same cell is missing, that subject
contributes no paired data to either hypothesis, so all paired
statistics below span the remaining four subjects. The pre-declared
pooled-with-degenerate and pooled-without-degenerate views therefore
coincide in composition for this dataset, a composition collapse we
state explicitly.

\subsubsection{4.2 H1 (replication): does the loop kill more than
one-shot?}\label{h1-replication-does-the-loop-kill-more-than-one-shot}

Pre-written support criterion (fixed before any run): H1 is supported if
and only if the pooled McNemar test comparing loop-same to oneshot
yields p \textless{} 0.05 and the direction favors loop-same (b
\textgreater{} c, where b counts mutants killed by loop-same only).

Per-subject paired 2x2, over each subject's baseline survivor set:

{\def\LTcaptype{none} 
\begin{longtable}[]{@{}
  >{\raggedright\arraybackslash}p{(\linewidth - 10\tabcolsep) * \real{0.1304}}
  >{\raggedleft\arraybackslash}p{(\linewidth - 10\tabcolsep) * \real{0.1739}}
  >{\raggedleft\arraybackslash}p{(\linewidth - 10\tabcolsep) * \real{0.1739}}
  >{\raggedleft\arraybackslash}p{(\linewidth - 10\tabcolsep) * \real{0.1739}}
  >{\raggedleft\arraybackslash}p{(\linewidth - 10\tabcolsep) * \real{0.1739}}
  >{\raggedleft\arraybackslash}p{(\linewidth - 10\tabcolsep) * \real{0.1739}}@{}}
\toprule\noalign{}
\begin{minipage}[b]{\linewidth}\raggedright
Subject
\end{minipage} & \begin{minipage}[b]{\linewidth}\raggedleft
both
\end{minipage} & \begin{minipage}[b]{\linewidth}\raggedleft
b (loop-same only)
\end{minipage} & \begin{minipage}[b]{\linewidth}\raggedleft
c (oneshot only)
\end{minipage} & \begin{minipage}[b]{\linewidth}\raggedleft
neither
\end{minipage} & \begin{minipage}[b]{\linewidth}\raggedleft
p (exact McNemar)
\end{minipage} \\
\midrule\noalign{}
\endhead
\bottomrule\noalign{}
\endlastfoot
graph-guard & 12 & 5 & 0 & 5 & 0.0625 \\
rag-guard & 21 & 4 & 0 & 0 & 0.125 \\
packaging & 36 & 32 & 0 & 1 & 4.66e-10 \\
idna & 123 & 64 & 0 & 0 & 1.08e-19 \\
\end{longtable}
}

Pooled (confirmatory view): b = 105, c = 0, p = 4.93e-32.

Under the pre-registered criterion this reads as supported, and we
report that for fidelity to the pre-registration; we then immediately
bound it. The pooled unit of analysis is the mutant, mutants cluster
within four subjects, and the arms are independently sampled pipelines
(Section 3.2), so the pooled inference is not valid for generalization
beyond these subjects and this comparison is descriptive: the loop-arm
pipeline outscored the one-shot pipeline on every subject, and across
303 paired baseline survivors not one mutant was killed by one-shot
generation but missed by the loop arm. The causal estimate of the loop's
incremental effect is Experiment 2's (Section 4.7), not this table's.
Two subjects individually clear p \textless{} 0.05 (packaging, idna) and
two do not (graph-guard at 0.0625, rag-guard at 0.125, both with small
discordant counts), but the direction is consistent in all four. The
pooled test treats mutants as independent when they in fact cluster by
subject, which inflates the nominal pooled significance; the pool is
also idna-heavy (64 of the 105 discordant kills) and packaging-heavy (32
of 105). We therefore report the per-subject tables alongside the pool
rather than resting on the pooled figure.

Notably, the instrument correction strengthened H1. Under the
defect-affected instrument the H1 split read b = 79, c = 5, with one
subject tied and one point against the loop. The truncation defect had
been silently removing critic rounds that loop-same should have run,
suppressing the loop's own kills. Correcting it moved packaging from a
4-to-4 tie to plus 32 and graph-guard from 0-versus-1-against to plus 5
(one accepted critic round of over 30,000 output tokens, impossible
under the old cap, killed 5 survivors). The corrected cells are new
stochastic draws, so no monotone counterfactual relation to the
truncated originals follows; we note only that the correction moved the
comparison further in the loop arm's favor.

\subsubsection{4.3 H2: does the cross-provider critic configuration kill
more? (pre-registered as
``cross-lineage'')}\label{h2-does-the-cross-provider-critic-configuration-kill-more-pre-registered-as-cross-lineage}

Pre-written support criterion: H2 is supported if and only if the pooled
McNemar comparing loop-cross to loop-same yields p \textless{} 0.05 and
the direction favors loop-cross. A null was pre-declared to be reported
with the same prominence as a positive result.

Per-subject paired 2x2:

{\def\LTcaptype{none} 
\begin{longtable}[]{@{}
  >{\raggedright\arraybackslash}p{(\linewidth - 10\tabcolsep) * \real{0.1304}}
  >{\raggedleft\arraybackslash}p{(\linewidth - 10\tabcolsep) * \real{0.1739}}
  >{\raggedleft\arraybackslash}p{(\linewidth - 10\tabcolsep) * \real{0.1739}}
  >{\raggedleft\arraybackslash}p{(\linewidth - 10\tabcolsep) * \real{0.1739}}
  >{\raggedleft\arraybackslash}p{(\linewidth - 10\tabcolsep) * \real{0.1739}}
  >{\raggedleft\arraybackslash}p{(\linewidth - 10\tabcolsep) * \real{0.1739}}@{}}
\toprule\noalign{}
\begin{minipage}[b]{\linewidth}\raggedright
Subject
\end{minipage} & \begin{minipage}[b]{\linewidth}\raggedleft
both
\end{minipage} & \begin{minipage}[b]{\linewidth}\raggedleft
b (loop-cross only)
\end{minipage} & \begin{minipage}[b]{\linewidth}\raggedleft
c (loop-same only)
\end{minipage} & \begin{minipage}[b]{\linewidth}\raggedleft
neither
\end{minipage} & \begin{minipage}[b]{\linewidth}\raggedleft
p (exact McNemar)
\end{minipage} \\
\midrule\noalign{}
\endhead
\bottomrule\noalign{}
\endlastfoot
graph-guard & 17 & 5 & 0 & 0 & 0.0625 \\
rag-guard & 25 & 0 & 0 & 0 & 1.0 (no discordant pairs, ceiling) \\
packaging & 68 & 0 & 0 & 1 & 1.0 (no discordant pairs) \\
idna & 187 & 0 & 0 & 0 & 1.0 (no discordant pairs, ceiling) \\
\end{longtable}
}

Pooled (confirmatory view): b = 5, c = 0, p = 0.0625.

Verdict: H2 NOT SUPPORTED. The pooled p of 0.0625 does not clear the
pre-registered 0.05 bar. The point direction favors loop-cross (all five
discordant pairs, from the single subject graph-guard, go the
cross-lineage way, and none go the other way), but the evidence does not
meet the pre-written criterion. This is an underpowered null, not a
demonstration of no effect: three of the four scorable subjects
saturated at 98.6 to 100 percent kill in both loop arms, leaving zero
discordant pairs, so the design had almost no room to detect a
cross-lineage difference even if one exists. The correct reading is no
effect detectable at this design's power, with a directionally positive
but non-significant residual on one subject.

\subsubsection{4.4 The central finding of Experiment 1: H2's earlier
signal was an instrument
artifact}\label{the-central-finding-of-experiment-1-h2s-earlier-signal-was-an-instrument-artifact}

Reported at the same prominence as any supported result, because it is
the paper's central contribution. The superseded analysis reported H2
SUPPORTED at a pooled p of 9.5e-66, with b = 217 and c = 0, an
apparently overwhelming cross-lineage effect. The corrected instrument
erased essentially all of it.

\begin{itemize}
\tightlist
\item
  packaging: b went from 32 to 0. The old loop-same cell's two critic
  rounds had both been silently truncated at exactly 16,000 output
  tokens each and rejected under unrelated-sounding notes, crediting
  zero kills. Rerun with truncation detection and a raised cap, the
  same-lineage critic's rounds were accepted and killed 32 more
  survivors, landing at 68 of 69, identical to loop-cross. The strongly
  supported reading is that the cap deleted the verbose model's rounds;
  because this rests on a single rerun of a cell class that carries real
  run-to-run variance, tester-round variance as a partial contributor
  cannot be fully excluded by the receipts, so this is the
  well-supported reading rather than a mechanical decomposition.
\item
  attrition-risk-ml: b went from 185 to excluded. Its old loop-same cell
  was likewise truncation invalidated, and the subject could not produce
  a valid replacement within the attempt budget, so the missing-cell
  rule removed it from the comparison entirely.
\item
  What remains is graph-guard's b = 5, a real gap between two accepted
  runs worth noting descriptively, and three subjects at or near ceiling
  with no discordant pairs at all.
\end{itemize}

In summary: the dramatic cross-lineage effect this experiment appeared
to measure was manufactured by a lineage-correlated instrument failure
and does not survive the instrument's correction. The output cap sat on
every call to the same-provider model, so it truncated the verbose
same-lineage critic and some tester rounds alike, never the terse
cross-lineage critic; the bias it injected into H2 specifically ran
through the same-lineage critic rounds, which is what manufactured the
apparent cross-lineage gap. The Discussion analyzes the mechanism, the
fail-loud repair, and why this generalizes to any cross-model comparison
run through an asymmetric harness.

\subsubsection{4.5 Cost per kill}\label{cost-per-kill}

Aggregated across counted cells with valid verdicts (total cost summed
and divided by total kills summed per arm, from full-precision receipts,
never averaged as a per-subject ratio):

{\def\LTcaptype{none} 
\begin{longtable}[]{@{}lrrr@{}}
\toprule\noalign{}
Arm & Total cost (\$) & Total killed & Aggregate \$/kill \\
\midrule\noalign{}
\endhead
\bottomrule\noalign{}
\endlastfoot
oneshot & 0.4844 & 254 & 0.0019 \\
loop-same & 2.4170 & 297 & 0.0081 \\
loop-cross & 1.3293 & 556 & 0.0024 \\
\end{longtable}
}

One asymmetry must be read alongside the 3.4-times gap between
loop-cross and loop-same: the oneshot and loop-cross aggregates each
span all five subjects, while loop-same spans only its four valid cells
(the missing attrition-risk-ml cell). loop-cross's 556 kills therefore
include attrition-risk-ml's 254, which loop-same never reached, so the
arms are aggregated over different subject sets and the ratio is not a
like-for-like per-subject comparison.

Because the arms aggregate over different subject sets, we draw no
comparative conclusion from this table. Experiment 2's like-for-like
accounting (Section 4.9) supersedes it: there, token receipts show that
output length and rejected rounds account arithmetically for most of the
billing difference between these two configurations, with no inference
about what capability- or budget-matched models would cost.

\subsubsection{4.6 Experiment 2: execution, and the confound
quantified}\label{experiment-2-execution-and-the-confound-quantified}

All twenty replicates completed with all three continuations validly
scored: twenty-one seeds were frozen (twenty plus one replacement,
below), no seed draw was rejected, and no cell is missing. The design's
premise is visible in the seeds themselves. Five independent round-0
draws per subject produced kill counts of 29, 32, 35, 36, and 45 (of 69
baseline survivors) on packaging, a sixteen-kill spread from sampling
alone, against a flat 12 (of 22) on every counted graph-guard draw and
127 to 132 (of 187) on idna: the initial-draw variance that Experiment
1's arms resampled is large and strongly subject-dependent, which is
precisely why its comparisons could not isolate the loop. Every
subject's pristine baseline measured identically on all of its draws,
and each seeded continuation re-verified its frozen measurement before
any Critic call.

The pre-registered contingency machinery fired twice, and both events
ran to completion without operator discretion. First, one continuation's
reproduction check failed: a mutant the frozen seed had measured as
surviving was measured as killed at injection time, a test
non-deterministic at the margin that had cleared the two-run flake check
(the disclosed residual class). The cell was receipted as invalid and
never scored, the replicate was invalidated wholesale, the seed was
retired in place, and the mandatory replacement draw and its three
continuations ran under the caps. Second, one continuation was killed
mid-run by a tooling timeout external to the instrument; it was
receipted as a crashed attempt and consumed the single pre-registered
rerun, which completed cleanly. Both events are logged in the deviations
file with their receipts.

\subsubsection{4.7 The causal estimate: the loop's incremental effect
over a frozen round
0}\label{the-causal-estimate-the-loops-incremental-effect-over-a-frozen-round-0}

Per subject, the mean incremental kills (Delta\_same) and incremental
kill rate (rho\_same) of same-lineage Critic rounds over the frozen
round 0, with 95\% bootstrap intervals over five replicates each; pooled
values are unweighted means of subject means with subject-cluster
bootstrap intervals:

{\def\LTcaptype{none} 
\begin{longtable}[]{@{}
  >{\raggedright\arraybackslash}p{(\linewidth - 4\tabcolsep) * \real{0.2727}}
  >{\raggedleft\arraybackslash}p{(\linewidth - 4\tabcolsep) * \real{0.3636}}
  >{\raggedleft\arraybackslash}p{(\linewidth - 4\tabcolsep) * \real{0.3636}}@{}}
\toprule\noalign{}
\begin{minipage}[b]{\linewidth}\raggedright
Subject
\end{minipage} & \begin{minipage}[b]{\linewidth}\raggedleft
mean Delta\_same {[}95\% bootstrap interval{]}
\end{minipage} & \begin{minipage}[b]{\linewidth}\raggedleft
mean rho\_same {[}95\% bootstrap interval{]}
\end{minipage} \\
\midrule\noalign{}
\endhead
\bottomrule\noalign{}
\endlastfoot
graph-guard & 5.0 {[}5.0, 5.0{]} & 0.50 {[}0.50, 0.50{]} \\
idna & 57.2 {[}55.0, 59.0{]} & 0.99 {[}0.97, 1.00{]} \\
packaging & 24.8 {[}11.8, 34.0{]} & 0.77 {[}0.39, 0.97{]} \\
rag-guard & 4.4 {[}3.2, 5.4{]} & 0.87 {[}0.77, 0.97{]} \\
\textbf{pooled} & \textbf{22.9 {[}4.7, 44.2{]}} & \textbf{0.783
{[}0.592, 0.935{]}} \\
\end{longtable}
}

Starting from an identical frozen suite, same-lineage Critic rounds
killed a mean 78\% of the survivors that suite left standing, and the
pooled interval's lower bound sits far from zero (under the stricter
two-stage bootstrap of Section 4.8, rho\_same's interval is {[}0.583,
0.964{]} and Delta\_same's is {[}4.65, 44.55{]}, both materially
unchanged). Because Delta is non-negative by construction, sign carries
no information and the pre-registered question was magnitude; the
magnitude is large on these subjects but heterogeneous: graph-guard's
loop killed exactly 5 of its 10 remaining survivors in every replicate
(a hard stop at its numeric-tolerance mutants), while idna's loop swept
nearly everything. The estimand this identifies is Section 3.10's narrow
one: the incremental kills of this guarded, add-only continuation,
conditional on these frozen seeds, subjects, models, and stopping rules.
Within that scope it is the causal counterpart of Experiment 1's H1,
isolated from the initial draw.

\subsubsection{4.8 The cross-configuration pilot: per-replicate values,
two readings, and
sensitivities}\label{the-cross-configuration-pilot-per-replicate-values-two-readings-and-sensitivities}

The within-replicate differences between the two critic configurations,
pre-registered as a pilot with no verdict attached. The complete
replicate-level values (this comparison's raw data):

{\def\LTcaptype{none} 
\begin{longtable}[]{@{}
  >{\raggedright\arraybackslash}p{(\linewidth - 8\tabcolsep) * \real{0.1765}}
  >{\raggedright\arraybackslash}p{(\linewidth - 8\tabcolsep) * \real{0.1765}}
  >{\raggedright\arraybackslash}p{(\linewidth - 8\tabcolsep) * \real{0.1765}}
  >{\raggedleft\arraybackslash}p{(\linewidth - 8\tabcolsep) * \real{0.2353}}
  >{\raggedleft\arraybackslash}p{(\linewidth - 8\tabcolsep) * \real{0.2353}}@{}}
\toprule\noalign{}
\begin{minipage}[b]{\linewidth}\raggedright
Subject
\end{minipage} & \begin{minipage}[b]{\linewidth}\raggedright
D per replicate (cross - same, kills)
\end{minipage} & \begin{minipage}[b]{\linewidth}\raggedright
rho\_gap per replicate
\end{minipage} & \begin{minipage}[b]{\linewidth}\raggedleft
subject mean D
\end{minipage} & \begin{minipage}[b]{\linewidth}\raggedleft
subject mean rho\_gap
\end{minipage} \\
\midrule\noalign{}
\endhead
\bottomrule\noalign{}
\endlastfoot
graph-guard & 5, 5, 5, 5, 2 & .50, .50, .50, .50, .20 & 4.4 & 0.440 \\
idna & 0, 0, 2, 1, 0 & 0, 0, .036, .017, 0 & 0.6 & 0.011 \\
packaging & 0, 0, 0, 39, 0 & 0, 0, 0, .975, 0 & 7.8 & 0.195 \\
rag-guard & 0, 1, 0, 0, 1 & 0, .167, 0, 0, .167 & 0.4 & 0.067 \\
\end{longtable}
}

Pooled (unweighted subject means, pre-registered cluster bootstrap): D =
3.3 {[}0.5, 6.1{]}; rho\_gap = 0.178 {[}0.039, 0.347{]}. Both pooled
intervals sit above zero. Under Experiment 1's independently-sampled
design the same comparison returned p = 0.0625 on five discordant pairs.

This pilot estimand is operational by pre-registration: every receipted
failure counts, so it estimates end-to-end deployability of the two
configurations under the specified harness, caps, and validators, not
critic capability. The replicate table shows why the distinction
matters, because the pooled gap decomposes into two mechanisms the
receipts separate. On graph-guard, the same-provider critic's rounds
were mostly accepted (16 of its 18 critic rounds across the five cells;
the two rejections were themselves 32,000-token truncations, in
replicates 1 and 2) and the arm plateaued at exactly 5 kills in every
replicate, including the three whose rounds were all accepted, against
the cross-provider critic's near-sweeps: evidence consistent with a
capability-related difference on that subject's numeric-tolerance
mutants, though we do not report per-arm generated-test volumes or token
budgets and therefore do not claim an identified capability effect. On
packaging, the entire subject mean rides on one replicate: the
same-provider critic's first round was rejected for a formatting
violation and its second hit the corrected instrument's raised output
cap at exactly 32,000 tokens (the third such truncation in Experiment 2,
and the only one that zeroed a cell's incremental kills), where the
truncation detector added by the autopsy's repair caught it, archived
the truncated reply, credited zero kills, and ended the cell at
Delta\_same = 0, while the cross-provider critic, from the identical
frozen state, killed 39 of the 40 remaining survivors. That single
replicate contributes 39 of the 66 total gap kills (59\%). The mechanism
that manufactured Experiment 1's false p = 9.5 x 10\^{}-66, verbose
replies meeting a mechanical output ceiling, is therefore a recurring
operational property of this model-harness pairing, and detection makes
it visible data rather than a laundered artifact, but detection does not
remove it from the operational estimate.

Because of that leverage, we report post-hoc sensitivities (labeled
exploratory; the pre-registered analysis is the pooled figure above).
Excluding the truncation replicate: D = 1.35, rho\_gap = 0.129.
Leave-one-subject-out (point estimates only; no intervals attached): D
ranges 1.8 (dropping packaging) to 4.3 (dropping rag-guard); rho\_gap
ranges 0.091 (dropping graph-guard) to 0.234 (dropping idna). Under a
stricter two-stage bootstrap that resamples subjects and replicates
within subjects, rho\_gap's interval is {[}0.017, 0.368{]} on the full
data and {[}0.004, 0.345{]} excluding the truncation replicate; D's is
{[}0.3, 9.0{]} and {[}0.1, 3.5{]}. The honest summary: the direction of
the configuration difference is stable under every sensitivity we ran,
never crossing zero, while its magnitude is dominated by a single
receipted operational failure and its interval, under the stricter
bootstrap, approaches zero. A capability-oriented reading should rest on
the graph-guard pattern (accepted rounds present in every cell, a
persistent kill gap) and treat the packaging replicate as deployability
evidence; neither reading supports a lineage inference, because the arms
bundle model, provider API behavior, and harness interaction by
construction.

Operationally, the cross-provider arm also failed less overall: 13 of
its 20 cells ended with every survivor killed against 5 of 20 for the
same-provider arm, with 3 rejected rounds against 10.

\subsubsection{4.9 Experiment 2 cost, like for
like}\label{experiment-2-cost-like-for-like}

Experiment 2's arms span identical subjects and replicates, so its cost
comparison has none of Experiment 1's denominator asymmetry. From
full-precision receipts, on explicit bases: the twenty accepted seeds
entering the analysis cost \$2.20 (all seed draws including the one
retired seed: \$2.33); the same-lineage arm \$11.14 for 457 incremental
kills (\$0.0244 per incremental kill); the cross-lineage arm \$2.01 for
523 incremental kills (\$0.0038 per incremental kill), 6.4x cheaper per
outcome at a similar rate card. The spend entering the analysis
(accepted seeds plus validly scored continuations) is \$15.34; the
all-receipted total is \$15.47; the difference is the retired seed draw
(\$0.13), the invalidated cell and the crashed attempt having billed
nothing before they stopped. The receipts decompose the difference
arithmetically: the same-lineage critic's rounds emitted 678,564 output
tokens against the cross-lineage critic's 100,545 (6.7x), and at the two
configurations' similar output rates that output alone accounts for
\$10.18 of the \$11.14 same-lineage arm cost and \$1.51 of the \$2.01
cross-lineage arm cost; the same-lineage critic also averaged 3.3 Critic
rounds per cell against 2.45, with ten rejected rounds against three.
Output volume and failed rounds, not per-token price, separate these two
configurations' costs; we make no claim about capability- or
budget-matched models.

All spend in one table, on explicit bases (each figure is derived in the
sections above):

{\def\LTcaptype{none} 
\begin{longtable}[]{@{}lrr@{}}
\toprule\noalign{}
Basis & Experiment 1 & Experiment 2 \\
\midrule\noalign{}
\endhead
\bottomrule\noalign{}
\endlastfoot
counted / analysis-entering & \$4.23 & \$15.34 \\
all-receipted (incl.~invalidated and rerun work) & \$11.05 & \$15.47 \\
disclosed unreceipted estimate (crash-billed calls) &
\textasciitilde\$0.63 & \$0 \\
\end{longtable}
}

\begin{center}\rule{0.5\linewidth}{0.5pt}\end{center}

\subsection{5. Discussion}\label{discussion}

\subsubsection{5.1 What the two experiments establish about the
loop}\label{what-the-two-experiments-establish-about-the-loop}

Experiment 1's H1 is a pipeline-level replication: the adversarial loop,
as a pipeline including its own initial draw, outscored one-shot
generation 105 discordant mutants to zero, confirming in a new setting
(agentic, repo-level, Python, with a mechanical mutation-kill oracle)
the direction established by MuTAP and AdverTest. It is not a causal
estimate of the loop, because each arm sampled its own round 0.
Experiment 2 supplies the causal version: with the round-0 suite frozen
and shared, the Critic loop's incremental effect is a within-replicate
paired quantity, and it is large, a mean 78\% of remaining survivors
killed (pooled rho\_same 0.783 {[}0.592, 0.935{]}), positive on nineteen
of twenty replicates (the twentieth is the truncation cell of Section
4.8), and heterogeneous across subjects in an informative way
(near-total on idna, a hard 50\% stop on graph-guard's numeric-tolerance
mutants). Replication in a new setting is a contribution but not a
discovery; the causal isolation, the per-outcome costs, and the fact
that the estimate survived an instrument-correction process and then a
design-correction process are what this pair of experiments adds. The
two experiments estimate related but distinct quantities under different
designs and budgets; Experiment 2's is the one its design identifies,
and it is the estimate this paper stands on.

\subsubsection{5.2 The instrument autopsy}\label{the-instrument-autopsy}

Section 4.4 reports the incident; this section interprets it. The
defect's essential property was not merely its size but its correlation
with the H2 arm. The ceiling applied to same-provider calls generally;
for H2, its consequential failures were verbose same-lineage Critic
rounds, so it preferentially deleted work from one arm. A
treatment-independent failure would not preferentially depress that arm;
this one could manufacture a difference -- and the manufactured result
was invisible from inside: a clean, well-formatted, enormous p-value,
with nothing in the number announcing that it measured the harness
rather than the models. When detection was added, the four affected
cells were invalidated; corrected-instrument replacements were run where
obtainable, with the one unresolvable cell scored missing, and the
apparent effect collapsed from p = 9.5 x 10\^{}-66 to p = 0.0625
(Section 4.4 gives the per-subject decomposition and the variance caveat
on the single-rerun cells).

\textbf{How it was caught.} Not by a runtime gate. This study caught
three earlier silent-corruption mechanisms mechanically, each before any
number reached a results file: a sandbox crash that would have laundered
into a false all-survived zero, and a scoping bug that would have
silently prevented freshly generated tests from ever being collected,
among them. The truncation defect slipped all of those gates, because it
produced no crash and no error, only a plausible rejection note. It was
caught by a final adversarial review that read the completed analysis,
after every counted cell had already run, and asked why exactly these
rounds were failing. That a defect this consequential survived every
mechanical check and was caught only by an independent skeptical read is
itself part of the finding.

\textbf{The fail-loud repair.} The fix makes the instrument halt loudly
rather than emit a plausible wrong number. The output ceiling is now a
single mechanical value read by both the request and a new check: every
reply's output-token count is compared against it, and a reply at the
ceiling raises a truncation error that is never retried (retrying a
capped call would bill again and likely truncate again). The round is
recorded honestly as ``truncated: output hit max\_tokens cap,'' its real
billed cost is metered, zero kills are credited, and the raw truncated
reply is archived as preserved evidence. A truncation at a cell's tester
round (round 0) fails that cell loudly as missing data rather than being
silently recovered; a truncated Critic round is recorded as rejected
with zero kills and the loop continues, which is exactly how Experiment
2's three recurrences were absorbed. The principle is fail-closed: when
the instrument cannot measure, it must say so, not hand back a confident
wrong number.

\textbf{A residual asymmetry we disclose rather than paper over.} The
truncation check exists only where a mechanical ceiling is declared,
which is the same-provider wrapper. The cross-lineage provider sends no
such ceiling and is therefore never truncation-checked. If that provider
ever truncated a reply against a server-side limit, this instrumentation
would launder it exactly the way the original defect did. We have no
evidence any such truncation occurred (every accepted cross-lineage
round is far from any plausible ceiling), but the detection asymmetry is
structural and we name it, because the entire point of this section is
that undisclosed asymmetries are how false results get made.

\subsubsection{5.3 The general lesson: cross-model comparisons can
inherit the asymmetries of their
harness}\label{the-general-lesson-cross-model-comparisons-can-inherit-the-asymmetries-of-their-harness}

The mechanical oracle at the core of this design was adopted
specifically to close the self-preference channel that undermines
LLM-as-judge evaluation, where a model scoring model output has no
ground truth. At the verdict level it did exactly that: no model ever
judged another model's output, and the kill verdict is a seeded defect
and a failing test. The autopsy's central point is that a different
bias, one level out from the verdict, survived the very mechanism meant
to make model-versus-model comparison safe. Closing the judgment channel
did not make the comparison neutral, because the apparatus that fed the
two models was not itself symmetric.

The recurrences in Experiment 2 (Section 4.8) sharpen the lesson: the
identical mechanism, a verbose model meeting a mechanical output
ceiling, fired three more times under the corrected instrument within
days of the repair, including on packaging, the same subject whose
truncated rounds had been Experiment 1's clearest illustration.
Asymmetries of this class are standing properties of a model-harness
pairing, not one-off bugs, and the difference between a false finding
and an honest data point is whether the harness detects the event or
launders it.

The lesson generalizes beyond the specific defect. A comparison is only
as neutral as the apparatus that runs it, and an apparatus that treats
the two models differently in any dimension correlated with the outcome
can produce a difference that looks like a model difference and is not.
Here the correlated dimension was output verbosity interacting with a
truncation ceiling; in another study it could be a timeout, a token
budget, a rate limit, a retry policy, or a parser that tolerates one
model's formatting quirks and not the other's. The defect is dangerous
precisely because it is invisible in the result: the manufactured
p-value was clean, well-formatted, and enormous. Nothing about the
number announced that it was measuring the harness rather than the
models.

Two practices follow, and this study argues for both by having needed
both. First, pre-registration of the hypotheses and success criteria, so
that a result cannot be quietly reinterpreted after the fact and so that
the difference between the pre-declared analysis and the corrected one
is itself a visible, dated record. Second, and less commonly practiced,
fail-closed instrumentation: a measurement pipeline should be built so
that when a component cannot do its job, it halts loudly instead of
emitting a plausible substitute. Most of the silent corruptions in this
study were caught because the instrument was progressively rebuilt to
fail loud; the one that reached the counted data was the one place the
instrument could still fail silently.

\subsubsection{5.4 What changed between the experiments, and what can be
attributed to
it}\label{what-changed-between-the-experiments-and-what-can-be-attributed-to-it}

Experiment 1's H2 and Experiment 2's cross-configuration pilot ask the
same question of the same subjects with the same models, and return
different answers: an inconclusive p = 0.0625 on five discordant pairs,
then a rate gap of 0.178 with a bootstrap interval above zero. Three
things changed together: the round-0 draw was frozen and shared
(removing initial-draw variance whose scale the seeds measured directly,
a sixteen-kill spread on packaging), replication went from one cell per
subject-arm to five, and counted-analysis spend increased 3.6-fold
(\$15.34 for the cells entering Experiment 2's analysis against \$4.23
for Experiment 1's counted cells; each experiment's additional
invalidated or replaced spend is receipted separately). The design does
not separate those contributions, so we do not attribute the changed
answer to variance removal alone; we claim only that the paired,
replicated design yields a more informative pilot estimate than the
original, and that Experiment 1's p = 0.0625 was an inconclusive reading
from a confounded design, not evidence of absence that Experiment 2
overturned.

The pilot framing still binds, for reasons fixed in the pre-registration
and sharpened by the data. Four subjects cannot power a configuration
claim, and the pooled interval, resampling four subject means, is honest
but coarse; under the stricter two-stage bootstrap it approaches zero.
The arms bundle model, provider API behavior, and harness interaction,
and Section 4.8 shows the bundle concretely: over half the measured gap
is one receipted operational failure of the same-provider configuration.
And the gap concentrates on the two subjects with headroom, so subject
selection materially shapes the pooled number. A confirmatory test needs
more and harder subjects and, ideally, multiple models per provider; the
frozen-round-0 design that would carry it now exists, is public, and is
validated.

\subsubsection{5.5 Why the autopsy and the review loop are the
contribution}\label{why-the-autopsy-and-the-review-loop-are-the-contribution}

Publishing the p = 9.5 x 10\^{}-66 result would have violated no
procedure: it was pre-registered, directionally as hypothesized, and
produced by real runs. Publishing H1 as a causal finding would have
violated none either; the first complete draft of this manuscript did,
until a cross-model review of the draft caught the round-0 confound its
authors and same-family reviewers had read past. The same practice
prevented both: the instrument, the completed analysis, and the
manuscript were each reviewed adversarially from outside the family that
produced them, and each level caught a defect everything below it had
missed -- the manuscript-level catch generating Experiment 2 itself.
Cross-model evaluation is now routine in software-engineering research
and practice; this study documents, with receipts, concrete ways such
evaluations silently go wrong, and what catching them costs and pays.

\begin{center}\rule{0.5\linewidth}{0.5pt}\end{center}

\subsection{6. Threats to Validity}\label{threats-to-validity}

\subsubsection{6.1 Construct validity}\label{construct-validity}

The oracle measures mutant kills, which is a proxy for fault detection,
not fault detection itself (the gap between mutation kills and
real-fault detection is the classic caveat of the mutation literature
{[}Papadakis et al.~2019; Just et al., ISSTA 2014{]}, and
pass-on-pristine plus mutant-kill is a partial answer to the oracle
problem {[}Barr et al.~2015{]}, not a resolution of it). Two gaps
follow. First, we perform no equivalent-mutant detection, so some
counted survivors may be semantically equivalent to the original and
unkillable by any test; every reported kill rate is therefore a lower
bound. This does not distort the paired comparisons, because both arms
in any comparison face the identical mutant set and an unkillable mutant
falls in the both-miss cell that McNemar discards, but it does mean
absolute kill rates should be read as floors. Second, we do not run
generated tests inside a syscall sandbox, so a sufficiently adversarial
generated test could in principle detect properties of the mutation
harness itself (timing, file paths, environment markers) and pass or
fail on that rather than on the mutant's behavior. Our guardrails
(add-only tests, pass-on-pristine validation, a flake check, and
wrong-oracle pruning) reduce this but do not mechanically close it. A
third construct threat concerns the cost measure: cost-per-kill does not
isolate lineage, because the same-lineage critic's cost is dominated by
its output verbosity rather than its lineage, so the cross-arm cost gap
reflects model output style as much as the critic mechanism.

\subsubsection{6.2 Internal validity}\label{internal-validity}

Experiment 1's comparisons carry the confound we disclose where the arms
are defined (Section 3.2): each arm independently regenerates its own
round-0 tester suite, so its loop-same and loop-cross cells differ in
both the tester draw and the critic lineage, and a single-run tester
draw carries substantial variance (the same tester produced round-0 kill
counts of 69, 168, 50, 62, and 65 across repeated attempts on one
subject; Experiment 2's seeds later measured a 29-to-45 spread on
packaging under identical prompts). Experiment 2 removes this confound
for the questions it re-measures, and its own internal-validity events
are disclosed in Section 4.6: one replicate was invalidated by the
reproduction check when a marginally non-deterministic test that had
cleared the two-run flake check measured differently at injection time,
and was replaced under the pre-registered mandatory flow; one cell
consumed its single permitted rerun after an external tooling timeout.
Both flows were outcome-blind by construction. A residual construct note
on the cross-lineage pilot: its measured gap includes the same-lineage
critic's operational failures (truncation, formatting rejections), which
we count deliberately as deployment-relevant but which a capability-only
reading would need to separate.

The study's central internal-validity event is the instrument artifact
of Sections 4.4 and 5.2. One residual remains: truncation is
mechanically detected only for the provider that declares an output cap,
so a server-side truncation on the other provider would still be
laundered. We have no evidence any occurred (every accepted
cross-lineage round is far from any plausible cap), but the detection
asymmetry is structural. A further, smaller asymmetry: one subject's
missing-cell rerun attempts ran with an added parallelism-control
configuration that its other two arms did not, because that
configuration was introduced to fight a deadlock; the kill oracle is a
deterministic pass or fail with no channel from a parallelism backend to
a verdict, but the instrument configuration across that subject's arms
is not byte-identical, and we record it. Finally, all five subjects are
public code that the models may have seen in pretraining; the
mutant-kill metric blunts but does not eliminate contamination, since a
memorized suite would still have to target the injected mutant to score
a kill.

\subsubsection{6.3 External validity}\label{external-validity}

The strongest limits are on generalization. One cell, the
same-lineage-loop arm of the largest subject (255 baseline survivors),
failed to produce a valid verdict within the pre-committed attempt
budget, and the fixed once-per-subject design has no backfill mechanism,
so that subject is dropped from both hypotheses entirely and the
pre-declared with-degenerate and without-degenerate pooled views
collapse to a single four-subject composition; the miss is disclosed,
not smoothed over. The remaining paired statistics therefore span four
subjects, of which three are the author's own public repositories, a
convenience sample; the study is single-language (Python) and holds a
single tester model fixed across all arms, so it does not speak to
whether a different tester lineage would change either result. Ceiling
effects further bound what H2 could show: three of the four scorable
subjects saturated at 98.6 to 100 percent kill in both loop arms,
leaving no headroom to detect a cross-lineage difference. Two
subject-specific notes also bear on generalization: one subject's target
module reads standard input in its command-line path, which mutation and
collection never exercise, and one subject's pinned module differs from
its own development-branch version, so any future run against that
branch would be a different subject, not a rerun.

\subsubsection{6.4 Statistical conclusion
validity}\label{statistical-conclusion-validity}

The pooled McNemar test treats individual mutants as independent units,
but mutants cluster by subject (one subject supplies 64 of the 105 H1
discordant pairs), and within-subject correlation is not modeled; this
inflates the nominal pooled significance, which is why per-subject
tables are pre-registered and reported alongside every pooled figure. No
a priori power analysis was performed, and the H2 result is an
underpowered null: with so few discordant pairs available, p = 0.0625 is
near the smallest two-sided value the design could produce, so the null
means no effect detectable at this power, never no effect. Kill counts
also carry real run-to-run variance from single-run cells, so
per-subject point estimates should be read as noisy; for example, one
subject's same-lineage-loop kills moved from 11 of 22 to 17 of 22
between its invalidated original run and its valid rerun. And the flake
check accepts a test after it passes on pristine code twice; a test that
is flaky roughly one third of the time still clears a two-run check
roughly four times in nine (flakiness in real suites is pervasive and
well-documented {[}Luo et al., FSE 2014{]}), so some residual flaky-kill
noise is expected and is not separately modeled; Experiment 2's
reproduction check caught exactly one such marginal test downstream, and
its seed was replaced under the pre-registered flow.

Experiment 2's statistics carry their own bounds, stated in the protocol
before its data existed. The pooled intervals cluster-bootstrap only
four subject means; with four clusters, percentile coverage is
unreliable and discrete, the pre-registered variant treats each subject
mean as fixed rather than propagating within-subject sampling variation,
and we therefore also report a post-hoc two-stage bootstrap (resampling
subjects, then replicates within subjects) and leave-one-out
sensitivities in Section 4.8, and use ``bootstrap interval'' rather than
``confidence interval'' throughout: these are descriptive uncertainty
statements for this subject set, not population-level inference. The
unweighted pooled mean is itself a descriptive average over two author
projects and two third-party modules, not an estimate of any naturally
defined population. Incremental kills are non-negative by construction
(tests are add-only from a shared frozen suite), so sign carries no
information and all conclusions are magnitude-with-interval statements.
Round-0 draw variance proved strongly subject-dependent (large on
packaging, nearly absent on graph-guard), so the flat five replicates
per subject buy unequal precision across subjects; the design fixed K in
advance rather than adapting it, and the heterogeneity is reported
rather than modeled.

\begin{center}\rule{0.5\linewidth}{0.5pt}\end{center}

\subsection{References}\label{references}

The inline markers of the form {[}arXiv:XXXX{]} in the text identify the
entries below.

\begin{itemize}
\tightlist
\item
  TestPilot: M. Schafer, S. Nadi, A. Eghbali, F. Tip. ``An Empirical
  Evaluation of Using Large Language Models for Automated Unit Test
  Generation.'' IEEE Transactions on Software Engineering, vol.~50 no.
  1, pp.~85-105, 2024. arXiv:2302.06527; DOI 10.1109/TSE.2023.3334955.
\item
  ChatTester: Z. Yuan, Y. Lou, M. Liu, S. Ding, K. Wang, Y. Chen, X.
  Peng. ``No More Manual Tests? Evaluating and Improving ChatGPT for
  Unit Test Generation.'' arXiv:2305.04207.
\item
  ChatUniTest: Y. Chen et al.~``ChatUniTest: A Framework for LLM-Based
  Test Generation.'' arXiv:2305.04764.
\item
  CoverUp: J. A. Pizzorno, E. D. Berger. ``CoverUp: Coverage-Guided
  LLM-Based Test Generation.'' arXiv:2403.16218; Proc. ACM Softw. Eng.
  (FSE 2025), DOI 10.1145/3729398.
\item
  CodaMosa: C. Lemieux, J. P. Inala, S. K. Lahiri, S. Sen.~``CodaMosa:
  Escaping Coverage Plateaus in Test Generation with Pre-trained Large
  Language Models.'' ICSE 2023.
\item
  Pynguin: S. Lukasczyk, G. Fraser. ``Pynguin: Automated Unit Test
  Generation for Python.'' ICSE-Companion 2022, pp.~168-172.
  arXiv:2202.05218; DOI 10.1145/3510454.3516829.
\item
  HITS: Z. Wang, K. Liu, G. Li, Z. Jin. ``HITS: High-coverage LLM-based
  Unit Test Generation via Method Slicing.'' ASE 2024. arXiv:2408.11324.
\item
  TestGen-LLM: N. Alshahwan et al.~``Automated Unit Test Improvement
  using Large Language Models at Meta.'' FSE 2024 (Industry).
  arXiv:2402.09171; DOI 10.1145/3663529.3663839.
\item
  MuTAP: A. M. Dakhel, A. Nikanjam, F. Khomh, M. C. Desmarais, H.
  Washizaki. arXiv:2308.16557; Information and Software Technology,
  2024, DOI 10.1016/j.infsof.2024.107468.
\item
  Meta ACH: C. Foster, A. Gulati, M. Harman, I. Harper, K. Mao, J.
  Ritchey, H. Robert, S. Sengupta. ``Mutation-Guided LLM-based Test
  Generation at Meta.'' FSE 2025 (Industry). arXiv:2501.12862.
\item
  LLMorpheus: F. Tip, J. Bell, M. Schafer. ``LLMorpheus: Mutation
  Testing using Large Language Models.'' arXiv:2404.09952.
\item
  Comprehensive Study: B. Wang, M. Chen, M. Deng, Y. Lin, M. Harman, M.
  Papadakis, J. M. Zhang. ``A Comprehensive Study on Large Language
  Models for Mutation Testing.'' arXiv:2406.09843; TOSEM.
\item
  AdverTest: Chang, Fang, Chen, Shi, Shen, Gu. ``Test vs Mutant:
  Adversarial LLM Agents for Robust Unit Test Generation.''
  arXiv:2602.08146.
\item
  A. Panickssery, S. R. Bowman, S. Feng. ``LLM Evaluators Recognize and
  Favor Their Own Generations.'' NeurIPS 2024. arXiv:2404.13076.
\item
  Y. Du, S. Li, A. Torralba, J. B. Tenenbaum, I. Mordatch. ``Improving
  Factuality and Reasoning in Language Models through Multiagent
  Debate.'' ICML 2024. arXiv:2305.14325.
\item
  Self-MoA: W. Li, Y. Lin, M. Xia, C. Jin. ``Rethinking
  Mixture-of-Agents: Is Mixing Different Large Language Models
  Beneficial?'' 2025. arXiv:2502.00674.
\item
  Refute-or-Promote: A. Agarwal. ``Refute-or-Promote: An Adversarial
  Stage-Gated Multi-Agent Review Methodology for High-Precision
  LLM-Assisted Defect Discovery.'' 2026. arXiv:2604.19049.
\item
  Great Models Think Alike: S. Goel, J. Struber, I. A. Auzina, K. K.
  Chandra, P. Kumaraguru, D. Kiela, A. Prabhu, M. Bethge, J. Geiping.
  ``Great Models Think Alike and this Undermines AI Oversight.'' 2025.
  arXiv:2502.04313.
\item
  TestForge: K. Jain, C. Le Goues. ``TestForge: Feedback-Driven, Agentic
  Test Suite Generation.''

  \begin{enumerate}
  \def\labelenumi{\arabic{enumi}.}
  \setcounter{enumi}{2024}
  \tightlist
  \item
    arXiv:2503.14713.
  \end{enumerate}
\item
  EvoSuite: G. Fraser, A. Arcuri. ``EvoSuite: Automatic Test Suite
  Generation for Object-Oriented Software.'' ESEC/FSE 2011, pp.~416-419.
  DOI 10.1145/2025113.2025179.
\item
  Registered Reports in Software Engineering: N. A. Ernst, M. T.
  Baldassarre. Empirical Software Engineering, 2023. arXiv:2302.03649.
\item
  Watts and Debts: A. S. Shany, S. Chandrasekar, K. Vaidhyanathan.
  ``Watts and Debts of Agentic Frameworks: An Empirical Study
  (Registered Report).'' ESEM 2026 (Registered Reports Track).
  arXiv:2606.10702.
\item
  Oracle Problem: E. T. Barr, M. Harman, P. McMinn, M. Shahbaz, S. Yoo.
  ``The Oracle Problem in Software Testing: A Survey.'' IEEE
  Transactions on Software Engineering, vol.~41 no. 5, pp.~507-525,
  2015. DOI 10.1109/TSE.2014.2372785.
\item
  Mutation Survey: M. Papadakis, M. Kintis, J. Zhang, Y. Jia, Y. Le
  Traon, M. Harman. ``Mutation Testing Advances: An Analysis and
  Survey.'' Advances in Computers, vol.~112, pp.~275-378, 2019. DOI
  10.1016/bs.adcom.2018.03.015.
\item
  LLM-as-judge: L. Zheng et al.~``Judging LLM-as-a-Judge with MT-Bench
  and Chatbot Arena.'' NeurIPS 2023 (Datasets and Benchmarks).
  arXiv:2306.05685.
\item
  Defects4J: R. Just, D. Jalali, M. D. Ernst. ``Defects4J: A Database of
  Existing Faults to Enable Controlled Testing Studies for Java
  Programs.'' ISSTA 2014, pp.~437-440. DOI 10.1145/2610384.2628055.
\item
  TOGA: E. Dinella, G. Ryan, T. Mytkowicz, S. K. Lahiri. ``TOGA: A
  Neural Method for Test Oracle Generation.'' ICSE 2022, pp.~2130-2141.
  arXiv:2109.09262; DOI 10.1145/3510003.3510141.
\item
  TitanFuzz: Y. Deng, C. S. Xia, H. Peng, C. Yang, L. Zhang. ``Large
  Language Models Are Zero-Shot Fuzzers: Fuzzing Deep-Learning Libraries
  via Large Language Models.'' ISSTA 2023, pp.~423-435.
  arXiv:2212.14834; DOI 10.1145/3597926.3598067.
\item
  Fuzz4All: C. S. Xia, M. Paltenghi, J. L. Tian, M. Pradel, L. Zhang.
  ``Fuzz4All: Universal Fuzzing with Large Language Models.'' ICSE 2024,
  Article 126. arXiv:2308.04748; DOI 10.1145/3597503.3639121.
\item
  Flaky Tests: Q. Luo, F. Hariri, L. Eloussi, D. Marinov. ``An Empirical
  Analysis of Flaky Tests.'' FSE 2014, pp.~643-653. DOI
  10.1145/2635868.2635920.
\end{itemize}

\end{document}